\documentclass{elsart}
\usepackage{epsfig}

\begin{document}
\begin{frontmatter}
\hyphenation{english}
\title{Tensor Glueball Photoproduction and Decay}
\author{Stephen R. Cotanch}
\address{Department of Physics, North Carolina State
University, Raleigh, NC  27695 USA}
\author{Robert A. Williams}
\address{  Nuclear  Physics Group, Hampton University,
Hampton, VA 23668 USA \\
and \\
Jefferson Lab,
12000 Jefferson Avenue, Newport News, VA 23606 USA }
\date{\today}
\maketitle
\begin{abstract}
Using vector meson dominance (VMD),
tensor glueball photoproduction cross sections, asymmetries 
and widths are calculated.  The predicted hadronic $V V'$ decays
are comparable for different  vector meson
($V = \rho , \omega$ and $\phi$)  channels with the $\omega \phi$ width the
largest but the  radiative $\omega \gamma$ and $\phi \gamma$ decays are 
suppressed  relative to $\rho
\gamma$  by over a factor of 2. This decay profile is distinct from typical meson
branching rates and may be a
useful glueball detection signature.  Results are compared to a previous 
VMD scalar glueball study.

\end{abstract}

\vspace{1.5cm}
\scriptsize{

}
\end{frontmatter}






Documenting hadron states with predominantly gluonic degrees of freedom,
i.e. glueballs,
has been a challenging and somewhat elusive pursuit.  Even though such
states are
consistent with quantum chromodynamics (QCD) and predicted by both
lattice simulations~\cite{bali,sexton,morningstar,lucini}
and gluonic models~\cite{ssjc,flsc}, clear experimental evidence
is lacking.
The purpose of this letter is to motivate additional experimental investigations by providing
estimates, based upon VMD, of tensor glueball photoproduction observables 
and also to detail a possible
decay signature for hadronic states with a significant gluonic
component. The latter entails comparable $VV'$ hadronic widths, with the largest 
branch to $\omega
\phi$ that promptly goes to $3 \pi K \bar{K}$, and somewhat suppressed 
$\omega \gamma$ and
$\phi \gamma$ radiative decays relative to $\rho \gamma$.
As discussed
below, this
decay signature is not expected for hadrons with a predominantly quark structure.
These results are  essentially model-independent since they follow
directly from the general principles of VMD, which has been found to agree  
with more fundamental
QCD based meson radiative calculations~\cite{scpm}, and the
flavor independence
of quark-gluonic couplings.    

Consider the radiative decay
$f_2 \rightarrow V(k') \gamma(k)$ of a neutral tensor hadron  with arbitrary quark,
gluon structure and mass $M_f$.  Here $k$, $k'$ are the momenta of the photon 
and vector meson with mass $M_V$ ($k'^2 = M_V^2$).
The most general, gauge invariant
$f_2V
\gamma$ vertex  is ~\cite{ren,ol}
\begin{eqnarray}
<\gamma (k) V(k') | f_2> &=&  \epsilon^\kappa \epsilon'^\lambda f^{\mu \nu}
A_{\kappa \lambda \mu \nu} (k, k') \ ,
\\
A_{\kappa \lambda \mu \nu} (k, k') &=& 4 \frac{g_1}{M_f^4} B_{\kappa \lambda \mu \nu} + 2
\frac{g_2}{M_f^2} C_{\kappa \lambda \mu \nu} \ ,
\\
B_{\kappa \lambda \mu \nu} (k, k') &=& (g_{\kappa \lambda} k \cdot k' -
k'_\kappa k_\lambda) k_\mu
k_\nu \ ,
\\
C_{\kappa \lambda \mu \nu} (k, k') &=& 2g_{\kappa \lambda} k_\mu
k_\nu + g_{ \lambda \mu}k'_\kappa
k_\nu  + g_{ \lambda \nu}k'_\kappa k_\mu - g_{ \kappa \mu}k_\lambda k_\nu -
g_{ \kappa \nu}k_\lambda k_\mu
\nonumber
\\ &-& \, \,  k \cdot k' (g_{ \kappa \mu}g_{ \lambda \nu} + g_{ \kappa
\nu}g_{ \lambda \mu}) \ ,
\end{eqnarray}
where $\epsilon^\kappa$, $\epsilon'^\lambda$ and $f^{\mu \nu}$ are the
photon, vector meson and $f_2$
polarization vectors and tensor, respectively.  Note that there are two
possible coupling
constants, $g_1$ and $g_2$,
which in VMD (also tensor meson dominance) are given by ~\cite{ren}
\begin{equation}
g_1 = 0, \, \, \; \; g_2 = e g_{f_2 V \gamma} = e  \sum_{ V'}\frac{g_{f_2 V
V'}} { f_{V'}} \ ,
\end{equation}
with $g_{f_2 V V'}$  the $f_2 V V'$ hadronic coupling constant, $f_{V'}$
the $V'$ leptonic
decay constant
and the sum is over all vector meson contributions consistent with isospin conservation for
the $f_2 V V'$
vertex. The radiative decay widths are  
\begin{equation}
\Gamma_{f_2 \rightarrow V \gamma} \;=\;
 \frac{2}{5} \;\alpha_e g_{f_2 V \gamma}^2 M_{f_2} ( 1 - x)^3[1 + \frac{x}{2} +
\frac{x^2}{6} ] \ ,
\label{vgwidth}
\end{equation}
and $\alpha_e = e^2/4\pi  = 1/137.036$,  $x =
M^2_V/M^2_{f_2}$. Focusing upon isoscalar  tensor hadrons ($I_{f_2} =   0)$ yields the
radiative couplings
\begin{eqnarray}
g_{f_2 \rho \gamma} =& \! \!\!\!\!\!\!\!\!\!\!\!\!
 \frac{g_{f_2 \rho \rho}} { f_{\rho}}, & \; \; \; \;  f_2 \rightarrow
\rho \gamma \ ,
\label{gvmdr} \\
g_{f_2 \omega \gamma} =&  \frac{g_{f_2 \omega \omega}} { f_{\omega}}+
\frac{g_{f_2
\omega \phi}} { f_{\phi}}, & \; \; \; \; f_2 \rightarrow \omega \gamma \ ,
\label{gvmdo} \\
g_{f_2 \phi \gamma} =&  \frac{g_{f_2 \phi \phi}} { f_{\phi}}+
\frac{g_{f_2  \phi
\omega}} { f_{\omega}}, &\; \; \; \; f_2 \rightarrow \phi \gamma 
\label{gvmdp} \ .
\end{eqnarray}
Since the $\rho$ and $\omega$ masses are almost equal ($M_{\rho^{0}} =$ 775.8 MeV,  
$M_{\omega} =$ 782.59 MeV), the ratio of  the $\omega$ to 
$\rho$ channel decays is simply
\begin{equation}
R_{\omega/\rho} = \frac{\Gamma_{f_2 \rightarrow \omega \gamma}}{\Gamma_{f_2 \rightarrow \rho
\gamma}}
= (\frac{g_{f_2 \omega \gamma}}{ g_{f_2
\rho
\gamma}})^2
\ .
\end{equation}
Application to tensor glueballs, i.e. $f_2 \rightarrow G_2$,
and assuming flavor independence for the glueball-vector meson couplings, 
$g_{G_2VV} = g_{G_2V'V''} $, yields
\begin{equation}
R_{\omega/\rho} = (\frac{f_{\rho}}{ f_{\omega}})^2 (1 + \frac{f_{\omega}}{ f_{\phi}})^2 \ .
\end{equation}
Hence the ratio of decay widths is entirely governed by the leptonic decay constants
whose magnitudes can be extracted  from  
$V \rightarrow e^+ e^-$ using
\begin{equation}
\Gamma_{V \rightarrow e^+ e^-} \;=\;
\frac{4 \pi \alpha_e^2}{3} \;  \frac{M_V}{f_V^2} \ . 
\end{equation}
The most recent measurements~\cite{pdg} yield  $|f_{\rho}|$ = 4.965, $|f_{\omega}|$ =
17.06 and $|f_{\phi}|$ = 13.38   for a relative reduction 
$R_{\omega/\rho} = 0.44$.   The 
$\phi \gamma
$ channel, which is also reduced by this factor, is further suppressed kinematically.  As
discussed below, suppression of radiative decays  to isoscalar vector meson channels is
not generally expected for tensor mesons 
since they will have different
$g_{f_2VV'}$ couplings reflecting their various flavor contents. 
Note also that the relative phase between the  
$\omega$ and $\phi$ couplings has been assumed to be the same as between their
respective decay constants. 
Depending upon phase convention (i. e. $\phi = \pm s {\bar s} $) the decay constants are
often cited with opposite phases in the literature (e.g. the  $SU(3)$
relation 
$f_\rho \sqrt{3} = - f_\omega sin (\theta)  = f_\phi cos (\theta)$~\cite{deswart} where
$\theta \approx 40^o$ is the $\omega
\phi$ mixing angle).  Consistency  requires the same relative sign between the couplings
$g_{G_2 \omega \omega}$ and $g_{G_2 \omega \phi}$ since the latter, like the $\phi$ decay
constant, is linear in the $\phi$ phase.  Because $f_\omega$ and $f_\phi$
are comparable in magnitude,  
$R_{\omega/\rho}$ is very sensitive to this relative phase and would be dramatically lower,
.0064, if indeed the net phase was negative.   It is
therefore important to more rigorously determine the relative phase of the vector meson
coupling and decay constants and further study is recommended.

Similarly, the scalar glueball radiative decay widths are~\cite{scbwprc}
\begin{equation}
\Gamma_{G_0\rightarrow V \gamma} \;=\;
\frac{ 1}{8} \; \alpha_e \; g^2_{G_0 V \gamma} \; \frac{M^3_{G_{0}}}{M_0^2} \; (1-x)^3
\label{G2gamma} \ ,
\end{equation}
where $g_{G_0 V \gamma}$ is also given in VMD by Eqs. (\ref{gvmdr}, \ref{gvmdo}, \ref {gvmdp}) 
with $f_2$ replaced by $G_0$,  $M_{G_0}$ is the scalar glueball mass,
$M_0$ is a reference mass fixed at 1 GeV and $x = M^2_V/M^2_{G_0}$.  Again VMD predicts the
same suppression factor for  radiative decays to
isoscalar vector meson channels.

The $G \rightarrow \gamma \gamma$ decays for $G_0$ and $G_2$ can also be obtained
from VMD 
\begin{equation}
\Gamma_{G_0\rightarrow \gamma \gamma} \;=\;
\frac{\pi }{4} \;\alpha_e^2 g_{G_0 V V}^2 
\left[ (\frac{1}{f_\rho})^2 + (\frac{1}{f_\omega} + \frac{1}{f_\phi})^2\right]^2
\frac{M_{G_0}^3}{M_0^2}
\label{G2gamma} \ , 
\end{equation}
\begin{equation}
\Gamma_{G_2\rightarrow \gamma \gamma} \;=\;
\frac{4\pi }{5} \;\alpha_e^2 g_{G_2 V V}^2\left[ (\frac{1}{f_\rho})^2 + (\frac{1}{f_\omega} +
\frac{1}{f_\phi})^2\right]^2M_{G_2} \ .
\end{equation}

To compute the radiative widths the hadronic couplings $g_{GVV}$ must be specified.  For
the scalar glueball, Ref.~\cite{scbwprc} uses the value $g_{G_0VV} = 3.43$,  but
there is an error in Eq.(37) of that paper which should instead
read, $g_{G_0 V V} [\frac{1}{f_{\omega}} + \frac{1}{f_{\phi}} ] = .62$,  yielding the 
slightly larger value $g_{G_0VV} = 4.65$.  The corrected coupling is now closer to 
4.23 which was obtained by an independent scalar glueball mixing analysis~\cite{bp}.
The tensor glueball coupling can be estimated  by assuming the hadronic
decays $G_2 \rightarrow VV'$ for $V=\rho$, $\omega$ and $\phi$ dominate 
and saturate the
entire tensor glueball width, i. e.
\begin{eqnarray}
\Gamma_{G_2} \;\approx \; \sum_{VV'} \Gamma_{{G_2} \rightarrow V V'} =
3 \Gamma_{G_2 \rightarrow \rho^0 \rho^0} + \Gamma_{{G_2} \rightarrow \omega
\omega} + \Gamma_{{G_2}
\rightarrow \phi \phi} + \Gamma_{{G_2} \rightarrow \omega \phi}
\ .
\label{fullwidth}
\end{eqnarray}
This of course represents more of an upper bound  for the coupling 
but for experimental planning it
should provide sufficient
photoproduction cross section estimates. 
Using Eq. (1) with the photon  replaced by a second vector meson,
$\gamma (k) \rightarrow V(k)$, the
tensor glueball hadronic widths are
\begin{eqnarray}
\Gamma_{G_2 \rightarrow V V'} \;&=&\;
S \frac{g_{G_2 VV}^2}{60 \pi} \; M_{G_2} ( 1 -2x_ +  + x^2_-)^{1/2}[6 - 9x_+ +
9x_+^2  \nonumber \\
&-&  (8 - x_+ - x_-^2) x_-^2 ]
\ ,
\label{vv'width}
\end{eqnarray}
where $S = 1/2$ if $V = V'$ and 1 otherwise, $x_\pm = x \pm x' $, $x
= M^2_V/M^2_{G_2}$ and $x' = M^2_{V'}/M^2_{G_2}$.  For identical mesons, $V = V'$ 

\begin{equation}
\Gamma_{{G_2} \rightarrow V V} \;=\;
 \frac{g_{{G_2}VV}^2}{20 \pi} \; M_{G_2} ( 1 - 4x)^{1/2}[1 -3 x + 6 x^2 ]
\ .
\label{vvwidth}
\end{equation}

Taking $f_2(2010)$ (mass 2011 MeV, total width 202 MeV) and
$f_2(2300)$ (mass 2297 MeV, total width 149 MeV) as  tensor glueball candidates,
Eqs. (\ref{fullwidth}, \ref{vv'width}, \ref{vvwidth}) yield, 
$g_{{G_2}VV} = 1.60$, for $f_2(2010)$ and a similar value,  $g_{{G_2}VV} =1.14$, for
$f_2(2300)$. Adopting the average, 1.37, for the tensor coupling,
the estimated glueball hadronic and radiative decays are summarized in Tables 1 
and 2, respectively.

\begin{table}[h]
\caption{Tensor  glueball hadronic decays in MeV}
\begin{center}
\begin{tabular}{|c||c|c|c|c|}
\hline
$V V' \rightarrow$ & $\rho^0 \rho^0$ & $\omega \omega$ & $\phi\phi$ & $\omega \phi$   \\
\hline
\hline
$\Gamma_{G_2(2010) \rightarrow V V'}$ & 26.2 & 25.8 & 10.3 & $\;33.0\;$  \\
$\Gamma_{G_2(2300) \rightarrow V V'}$   & 37.2 & 36.8 & 20.3 &  $\;44.7\;$  \\
\hline
\end{tabular}
\end{center}
\end{table}

\begin{table}[h]
\caption{VMD  glueball electromagnetic decays in keV}
\begin{center}
\begin{tabular}{|c||c|c|c|c|}
\hline
$V \rightarrow$ & $\rho$ & $\omega$ & $\phi$ & $\gamma$   \\
\hline
\hline
$\Gamma_{G(1700) \rightarrow V \gamma}$   & $\;1950\;$ & $\;844\;$  &
$\;453\;$ & $\;15.1\;$  \\
\hline
\hline
$\Gamma_{G_2(2010) \rightarrow V \gamma}$  & 298 & 129 & 91.6 & $\;1.72\;$  \\
$\Gamma_{G_2(2300) \rightarrow V \gamma}$   & 377 & 164 & 128 &  $\;1.96\;$  \\
\hline
\end{tabular}
\end{center}
\end{table}

From the Tables it is clear that VMD and flavor independence predict roughly
comparable hadronic $VV'$ widths.
It is interesting that the largest branch is to the $\omega \phi$ channel which has a clear,
novel $3 \pi K \bar{K}$ prompt decay. Also noteworthy are the suppressed $\omega \gamma$ and
$\phi
\gamma$ decays relative to $\rho \gamma$. Hence even though the gluonic coupling has been assumed
flavor blind, consistent with QCD, the glueball widths are not.  As mentioned above this decay
signature is not expected for mesons and several published studies find no
$\omega$/$\rho$ suppression in meson radiative decays.  Indeed Ref.~\cite{db}, which also uses VMD
for scalar meson radiative decays, actually
predicts an enhancement for $R_{\omega/\rho}$  by an order of magnitude.
Related,   tensor meson decay calculations~\cite{pk} to vector and pseudoscalar meson
channels reveal branching ratios that are very
sensitive to flavor, varying by over an order of magnitude. Further,
a recent meson decay model study~\cite{cdk}, which compliments this work by advocating
radiative decays as a flavor filter to clarify glueball mixing,  predicts
extremely flavor dependent radiative decays of scalar mesons. That investigation incorporates 
decay width renormalizations 
from mixing (see below) with a glueball component~\cite{ck} but does not include contributions
from glueball decays.  Similarly, Ref.~\cite{ji} repeats that analysis,
again not including direct glueball decays,
with relativistic quark model corrections and finds the same decay pattern but all widths are
reduced by 50 to 70\%.  Both studies detailed marked sensitivity of $f_0 \rightarrow \rho \gamma$
and $\phi \gamma$ decay widths to mixing
and quark flavor.

Concerning experimental evidence for $2^{++}$ isoscalar  hadrons with mass near 2 GeV,
the most recent PDG report~\cite{pdg} list six states: $f_2(1910)$, $f_2(1950)$,
$f_2(2010)$, $f_2(2150)$, $f_2(2300)$ and
$f_2(2340)$.  Also there is the $f_J(2220)$ which is a tensor candidate
but it, along with the
$f_2(1910)$ and
$f_2(2150)$, is omitted from the more important PDG summary table.  For the four firm 
tensor states there
is limited decay data and no quantitative branching
ratios.  The specific observed decays are: $\phi \phi$ for $f_2(2010)$ and
$f_2(2340)$;  $\phi \phi$, $K \bar{K}$ and $\gamma \gamma$ for  $f_2(2300)$;
$K^*(892) \bar{K}^*(892)$, $\pi^+ \pi^-$,  $4\pi$, $\eta \eta$,
$K \bar{K}$ and $\gamma \gamma$ for $f_2(1950)$.  There is a clear need for additional, 
more detailed measurements.

A final comment about glueball decays is in order regarding
quarkonia-glueball mixing.  In addition to the investigations discussed above, there have 
been several other
mixing studies involving scalar hadrons~\cite{bp,ac,lw,gf} but no published worked treating tensor
states in the 2 GeV region which is the focus here. For all theoretical models,
the isoscalar $2^{++}$ $q
\bar{q}$ states calculated in this mass region will mix with predicted nearby 
tensor glueballs and
this will alter the unmixed decay scheme. If the mixing is weak
the predicted VMD decay profile
will not be appreciably modified and
may be effective in identifying the existence of glueball
dominated states.  For strong or maximal mixing,
the branching ratios will depend upon model details and the  VMD predictions  will  be
affected  by hadronic couplings
in the quark sector, especially their flavor dependence.  In general significant
mixing will   distort the simple VMD glueball decay signature of  
suppressed $\omega \gamma$ and $\phi \gamma$ but comparable
$V V'$ decay rates. 
An improved mixing analysis, including decay  contributions
from both the quark~\cite{flsc2} and glue sectors, is in progress and will be reported
in a future communication.

The glueball couplings can also be used to describe the photoproduction process,
$\gamma(k,\lambda) + p(p,\sigma)  \rightarrow   
G(q, \lambda') + p(p',\sigma')  $,
where the energy-momentum 4-vectors (helicities) for the photon,
proton, glueball and  recoil proton 
are  $k \; (\lambda$), $p$ ($\sigma$), $q \; (\lambda')$ and  $p'$ ($\sigma'$),
respectively.
In the helicity representation the scalar glueball photoproduction  amplitude,
$<G_0 \;p\;|
\;\hat{T}\; |\;\gamma
\; p>$, is
\begin{equation}
<G_0 \;p\;| \;\hat{T}\; |\;\gamma \; p> =
 \; \epsilon_{\mu}(\lambda) \;
{\mathcal H}^{\mu}_{\sigma' \sigma} \equiv \epsilon \cdot {\mathcal H} \ ,
\end{equation}
with $\epsilon_{\mu} (\lambda)$ 
the  photon
polarization 4-vector  and
${\mathcal H}^{\mu}_{\sigma' \sigma}$ the hadronic current
obtained by application of Feynman rules to the tree level $s = (k + p)^2$,
$t = (q - k)^2$ and
$u = (p' - k)^2$ channel QHD diagrams.
The spin-averaged scalar glueball photoproduction cross section is
\begin{eqnarray}
\frac{d\sigma^{G_0}}{dt} = \frac{\pi}{4\omega_{cm}^2} \sum_{ \lambda \sigma' \sigma } 
|<G_0 \, p| \hat{T} |\gamma \,  p>|^2 = \frac{\pi}{4\omega_{cm}^2} \;
\sum_{ \sigma' \sigma }
[|\; {\mathcal H}^{1}_{\sigma' \sigma} \;|^2 + |\; {\mathcal H}^{2}_{\sigma'
\sigma} \;|^2]  \ , \nonumber
\end{eqnarray}
where $\omega_{cm}$ is the photon $cm$ energy.
As detailed in Ref.~\cite{scbwprc}, the glueball
cross section is dominated by $t$ channel exchanges for $\theta_{cm} < 65^0$.  Accordingly
only the $t$ channel amplitude is calculated 
and  from Ref.~\cite{scbwprc} this  is
\begin{eqnarray}
{\mathcal H}^{\mu}_{\sigma' \sigma} &=& \sum_{V= \rho, \omega, \phi} 
 \frac{e g_{{G_0}V\gamma}}{M_{0}}g_{VNN}
   F_{t}(t)  \Pi_{V}(t)
    \bar{u}(p',\sigma')  [ \gamma^{\mu}  + 
i  \frac{\kappa_V^T}{M_0} \; \sigma^{\mu \alpha} k'_{\alpha} ]
u(p,\sigma) \ , \nonumber
\label{eqt}
\end{eqnarray}
with $k' = p' - p$, $t = k'^2$.
The hadronic
form factor, $F_{t}(t) $, vector meson propagator, $\Pi_{V}(t)$, and remaining
vector-nucleon couplings and transition moments are specified in Ref.~\cite{scbwprc}.

Because of higher spin, the tensor glueball production amplitude is more complicated.
Invoking vector and tensor dominance and Eq. (1), the photoproduction amplitude is
\begin{eqnarray}
<G_2 \;p\;| \;\hat{T}\; |\;\gamma \; p> &=& \frac{2 f^{\mu \nu}}{M^2_{G_2}} 
[2 \epsilon \cdot \bar{{\mathcal H}} k_\mu k_\nu - k \cdot \bar{{\mathcal H}} (\epsilon_\mu k_\nu +
\epsilon_\nu k_\mu) \nonumber \\
&-& k \cdot k' (\epsilon_\mu \bar{{\mathcal H}}_\nu + \epsilon_\nu \bar{{\mathcal H}}_\mu)] \ .
\end{eqnarray}
The
hadronic current $\bar{{\mathcal H}}$ has the same form as the scalar glueball result,
${\mathcal H}$, except that the ratio $\frac{ g_{{G_0}V\gamma}}{M_{0}}$ is replaced
by $\frac{ g_{{G_2}V\gamma}}{M_{G_2}}$.
Again focusing on forward angles, only  $t$ channel diagrams are evaluated and 
since the formulation is covariant, the Gottfried-Jackson or glueball rest frame
is used for mathematical convenience.  The spin-averaged
tensor glueball production cross section is

\begin{eqnarray}
\frac{d\sigma^{G_2}}{dt} \;=\; 
\frac{\pi}{4\omega_{cm}^2} \;
\sum_{ \sigma' \sigma } \left[a (|\; \bar{{\mathcal H}}^{1}_{\sigma' \sigma} \;|^2 
+ |\; \bar{{\mathcal H}}^{2}_{\sigma' \sigma} \;|^2) +
b|\; \bar{{\mathcal H}}^{0}_{\sigma' \sigma} + (\frac{1 + y} {1 -y}) 
\bar{{\mathcal H}}^{3}_{\sigma' \sigma}\;|^2 \right]  \ , \nonumber
\end{eqnarray}
where $a = 4(1-y)^2(1+y^2/6)$, $b = (1-y)^4$, $y = t/M_{G_2}^2$ 
and $M_{G_2} =$ 2.011
GeV is the mass used in the cross section predictions presented here.  

For the above specified glueball couplings, the  tensor and scalar
photoproduction cross sections  are displayed in Figs. 1, 2 and 3.  Figures 1 and 2 depict the lab
energy  dependence for the forward $cm$ angles  $\theta_{cm} = 0^o$ and $25^o$, respectively,
while Fig. 3 shows the angular distribution for 6 GeV photon lab energy.   
Again, since only $t$ channel amplitudes are included, results for angles greater than 
$60^o$ should be ignored.  In contrast to the radiative widths, the cross sections are insensitive
to the relative  phase between the $\omega$ and $\phi$ couplings since $\rho$ exchange dominates (see
Eq.(7)).  While the scalar glueball cross section is somewhat larger, it is noteworthy that the
magnitude of both cross sections  is sufficient to expect reasonable count rates. Indeed, measurements
of this process, including vector meson decays, would appear feasible for the evisioned Hall D project
at Jefferson Lab.

Finally, photon transverse asymmetry observables, $A_{\gamma \bot}$,
are also predicted and are displayed in Figs. 4 and 5 for the respective $cm$ angles
$0^o$ and $25^o$.  The scalar glueball asymmetry is greater than the tensor and both 
are large and increase with energy.

In summary, both tensor and scalar glueball cross sections, asymmetries and decay observables 
have been predicted using VMD and flavor independence.  The results indicate
that photoproduction cross sections are measurable and
that by detecting comparable hadronic $VV'$ decays, especially the novel
$\omega \phi \rightarrow 3 \pi K \bar{K}$ branch, in conjunction with suppressed
$\omega \gamma$, $\phi \gamma$ transitions relative to $\rho \gamma$,
it may be possible to identify states having a significant
gluonic component.

\section*{Acknowledgements}

This work was supported by  DOE Grant No.DE-FG02-97ER41048.



\newpage


\begin{figure}
\psfig{figure=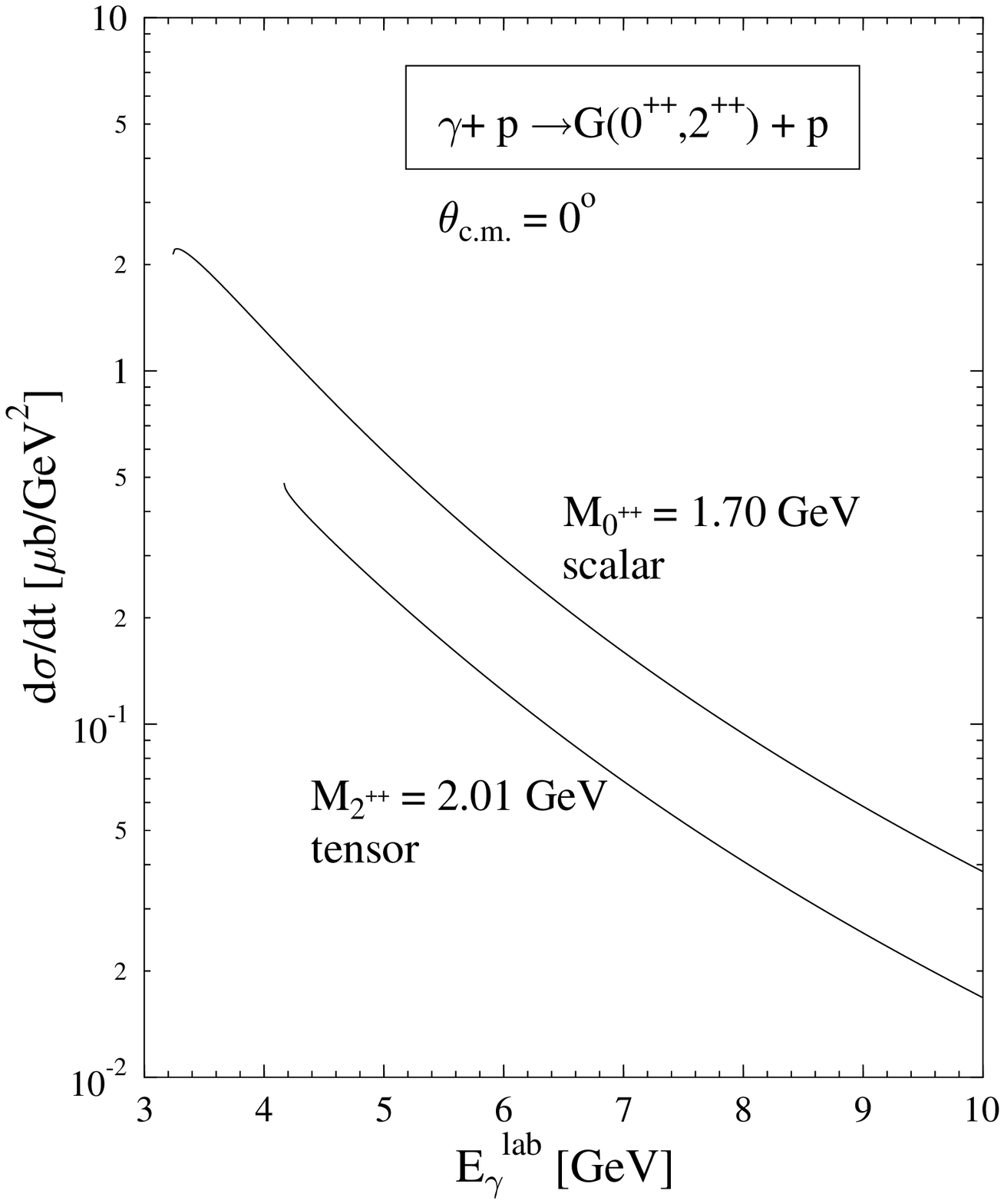,width=5.1in,height=6.6in}
\caption{Predicted  scalar and tensor glueball  cross sections vs lab
energy for ${\theta}_{cm} = 0^o$.}
\end{figure}

\begin{figure}
\psfig{figure=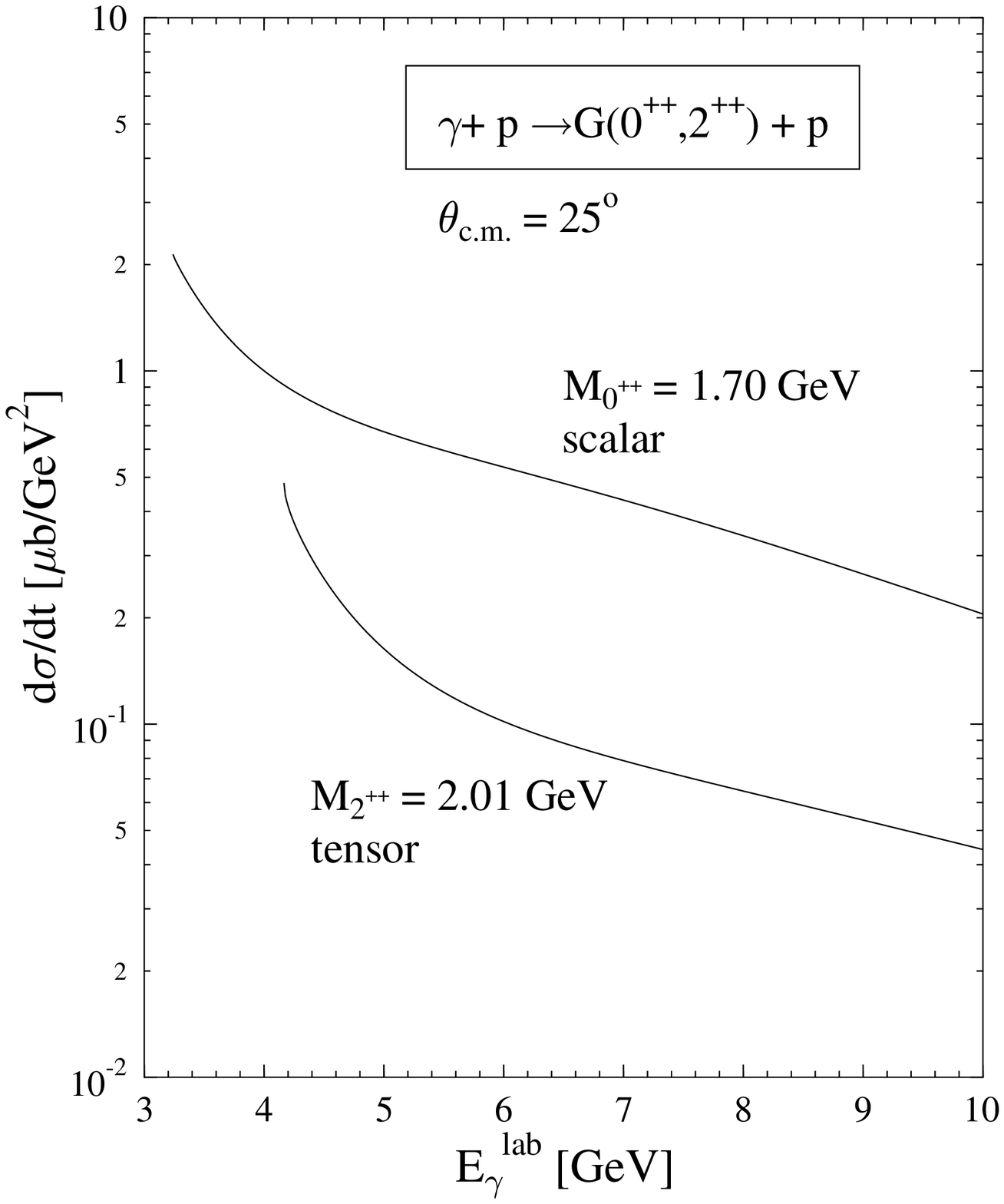,width=5.1in,height=6.6in}
\caption{Predicted scalar and tensor glueball  cross sections vs lab
energy for ${\theta}_{cm} = 25^o$.}
\end{figure}

\begin{figure}
\psfig{figure=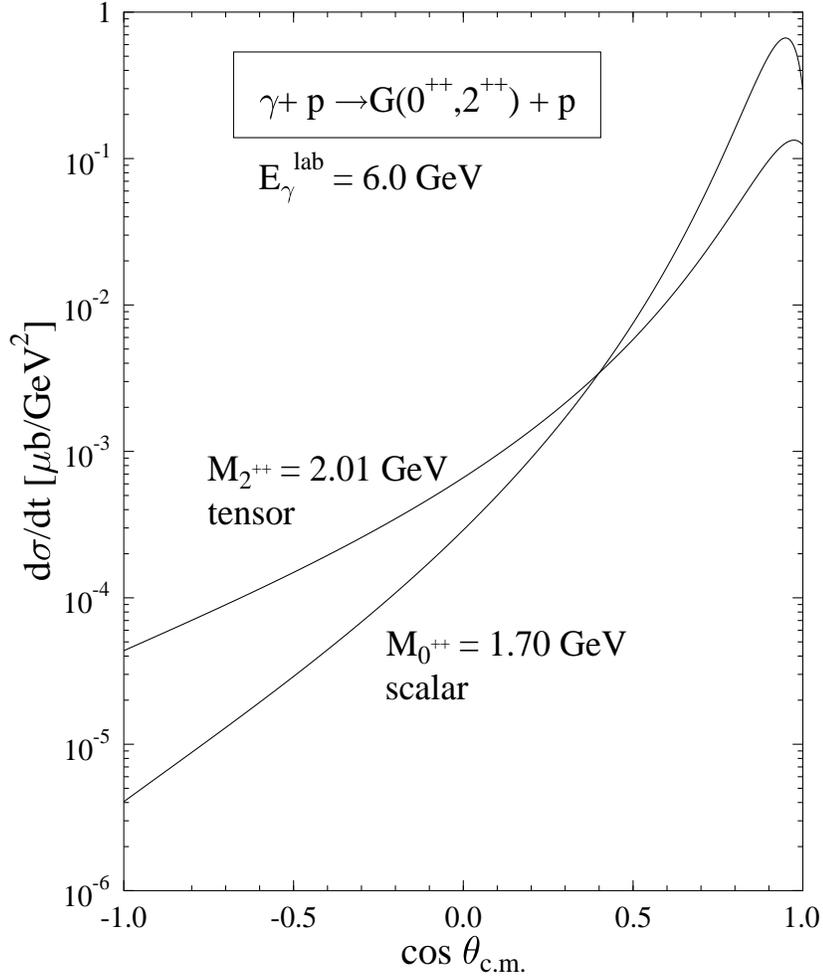,width=5.1in,height=6.6in}
\caption{Predicted scalar and tensor glueball  cross sections vs $cm$
angle.}
\end{figure}

\begin{figure}
\psfig{figure=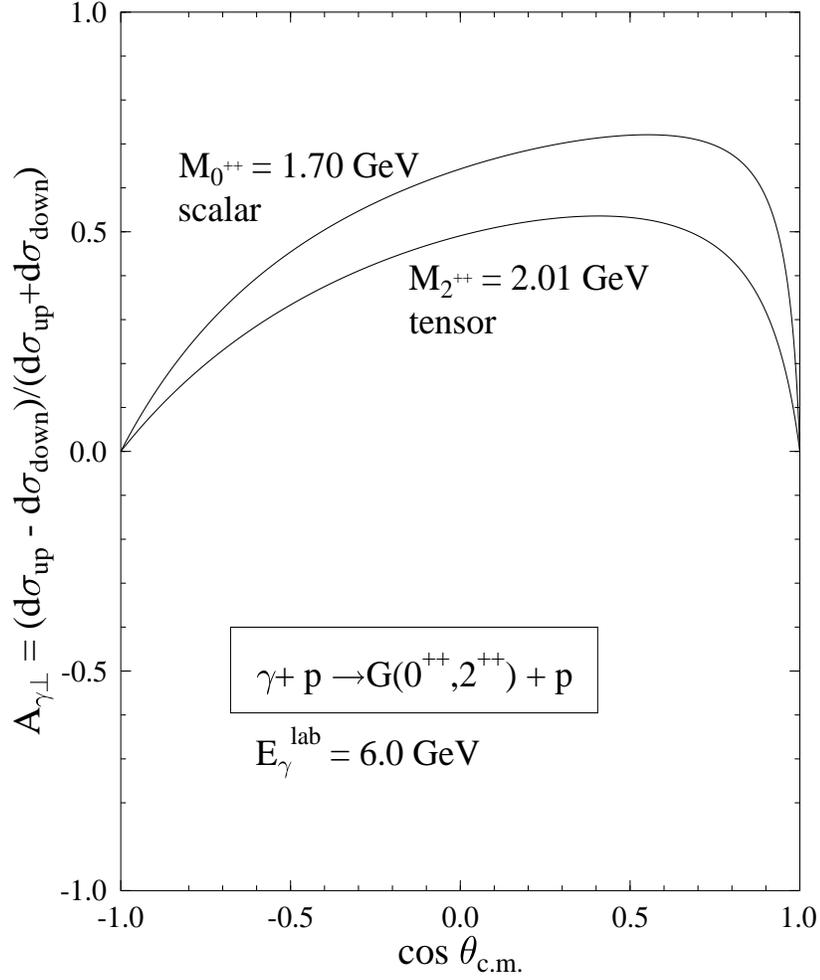,width=5.1in,height=6.6in}
\caption{Predicted  scalar and tensor glueball  transverse asymmetry vs $cm$
angle.}
\end{figure}

\begin{figure}
\psfig{figure=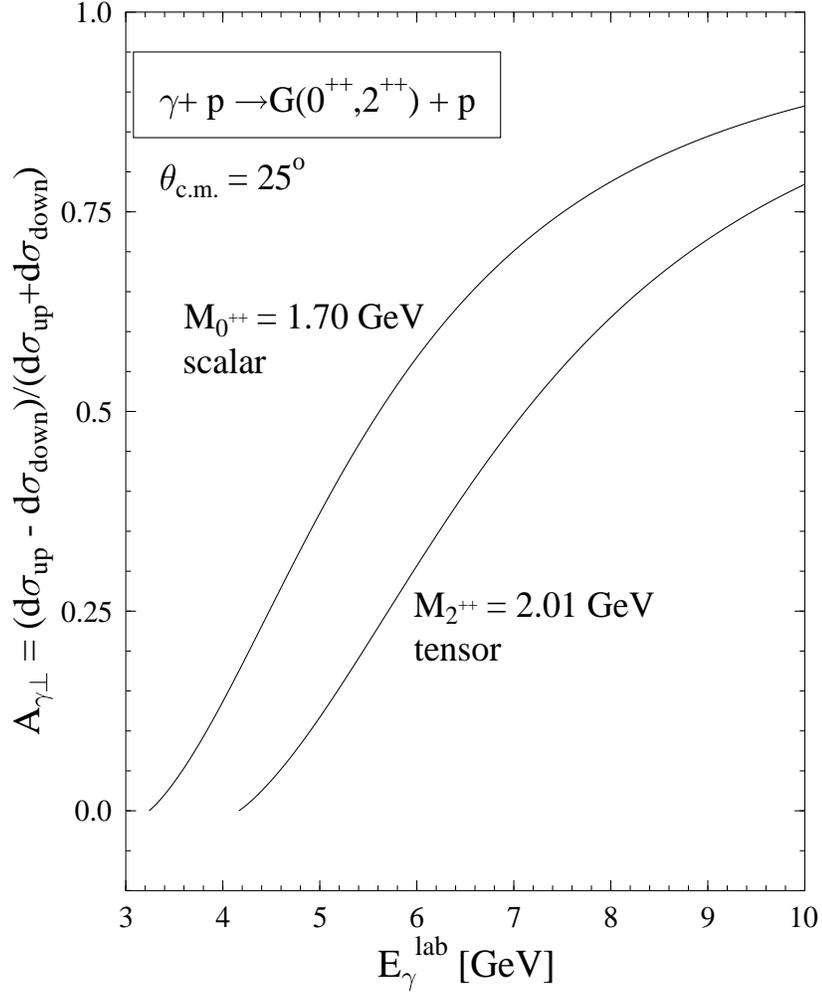,width=5.1in,height=6.6in}
\caption{Predicted  scalar and tensor glueball  transverse asymmetry vs lab
energy.}

\end{figure}

\end{document}